\documentstyle[aps,pra,multicol,amssymb,epsf,float]{revtex}

\newcommand{\ket}[1]{|{#1}\rangle }

\newcommand{\ketbra}[2]{|{#1} \rangle \langle  {#2} |}

\title{Quantum telecloning and multiparticle entanglement} 
\author{M. Murao$^1$, D. Jonathan$^1$, M. B. Plenio$^1$ and V. Vedral$^2$}
\address{ $^1$Optics Section, The Blackett Laboratory, Imperial College,
London SW7 2BZ, United Kingdom\\ 
$^2$ Centre for Quantum Computing,
Clarendon Laboratory, University of Oxford, Parks Road, Oxford OX1
3PU, United Kingdom} 
\date{\today}

\begin{document}
\draft
\maketitle
\begin{abstract}
A quantum telecloning process combining quantum teleportation and
optimal quantum cloning from one input to $M$ outputs is presented.
The scheme relies on the establishment of particular multiparticle
entangled states, which function as multiuser quantum information
channels.  The entanglement structure of these states is analyzed and
shown to be crucial for this type of information processing.
\end{abstract}

\pacs{PACS numbers: 03.67.-a, 03.67.Hk, 89.70.+c}

\begin{multicols}{2}

\section{Introduction}
\label{intro}

Quantum information-processing systems display many features which are
unknown in the classical world. Well-known examples include
teleportation \cite{Bennett0}, superdense coding \cite{Bennett1} and
the ability to support novel cryptographic and computational protocols
\cite{Bennett2,Ekert1}. Central to many of these applications is the
existence of entanglement between a {\it pair} of distant quantum
systems \cite{Plenio}. For instance, in the case of teleportation, the
establishment of a maximally entangled state of two distant qubits
allows an arbitrary unknown 1-qubit state to be conveyed from one
distant party to another with perfect fidelity.

The consequences of {\it multiparticle} entanglement involving several
distant parties have not yet been explored as extensively. An early
application was the use of Greenberger-Horne-Zeilinger (GHZ) states to
provide inequality-free tests of quantum mechanics versus local
hidden-variable theories \cite{Greenberger}. More recently,
multiparticle correlations have been shown to decrease the
communication complexity of certain multiparty calculations (i.e., to
reduce the amount of communication needed to realize a computation
involving data from several distant parties) \cite{Cleve}. Recent
developments also include state purification protocols for
multiparticle systems \cite{Murao}, schemes for basic manipulation of
multiparticle states via entanglement swapping \cite{Bose}, and
quantum secret sharing \cite{Hillery}.

Another important application of multiparticle entanglement is in
distributed quantum computing \cite{Ekert2}, where several distant
parties (Alice, Bob, Claire, etc.) share an initial entangled state,
and are asked to perform a given computational task using only local
operations and classical communication. The problem is to
find a protocol that completes the task using the least possible
resources (in particular, the minimum amount of initial nonlocal
entanglement, which is an `expensive' resource).

In this paper, we investigate this problem for the following scenario:
Alice holds an unknown 1-qubit quantum state $\left \vert \phi \right \rangle$
and wishes to transmit identical copies of it to $M$ associates (Bob,
Claire, etc.). Of course, the quantum no-cloning theorem \cite{Wootters} 
implies that these copies cannot be {\it perfect}. The best Alice can do is
to send {\it optimal quantum clones} of her state (the most faithful copies
allowed by quantum mechanics \cite{Buzek0,Buzek1,Gisin,Buzek2,Bruss}; see also 
section \ref{sec:uqcm}), which we assume are sufficient for her purposes. 
The computational task Alice must perform is therefore to generate $M$ 
optimal quantum clones of a 1-qubit input and distribute them among distant
parties.

The most straightforward protocol available to Alice would be to
generate the optimal clones locally using an appropriate quantum
network \cite{Buzek1,Buzek2} and then teleport each one to its
recipient by means of previously shared maximally entangled
pairs. This would require $M$ units of initial entanglement (e-bits),
as well as the sending of $M$ independent 2-bit classical messages
(one for each measurement result). It would also require Alice to run
a computationally expensive local network involving several extra
qubits and 2-qubit operations. In contrast, as we shall see ahead, far
cheaper strategies can be found (requiring only $O\left(\log_2
M\right)$ e-bits), provided Alice and her associates share particular
multiparticle entangled states. In this case, it is possible to
simultaneously convey all $M$ copies by means of a {\it single}
measurement on Alice's qubit. Alice only needs to publicly broadcast
the $2$ bits which determine her measurement result , after which each
recipient performs an appropriate local rotation conditioned on this
information. This `telecloning' is reminiscent of the well-known
teleportation protocol of Bennett {\it et al} \cite{Bennett0}. Indeed,
it can be seen as the natural generalization of teleportation to the
many-recipient case.

At this point, we should note that a similar proposal for telecloning
$M=2$ copies has been put forward by {Bru\ss} {\it et al}
\cite{Bruss}. In their case, however, the procedure was not directly
scalable to $M>2$. Moreover, the protocol was somewhat awkward,
involving the deliberate discarding of information. Our scheme avoids
both drawbacks, generating any number of copies and involving only 
classical communication, local unitary rotations and one local measurement.

Our work is organized as follows: in section \ref{sec:pre} we give a
summary of relevant results concerning teleportation and optimal
universal quantum cloning. In section \ref{sec:tc} we present our
telecloning protocol; section \ref{sec:ent} is devoted to analyzing 
the entanglement properties of the multiparticle telecloning states. 
Open questions raised by our study are discussed in section \ref{sec:open}. 
Finally, in section \ref{sec:summary} we present our conclusions.

\section {Preliminaries}
\label{sec:pre}
\subsection {Teleportation}
The teleportation protocol \cite{Bennett0} allows an unknown state
$\left \vert \phi \right \rangle_X$ of a quantum system $X$ to be
faithfully transmitted between two spatially separated parties (Alice
and Bob). The essential steps of this procedure (say in the simplest
case where $X$ is a 1-qubit system) are as follows: first and
foremost, Alice and Bob must share a maximally entangled state of two
qubits A and B, such as $\left \vert \Phi^{+} \right \rangle =
\frac{1}{\sqrt{2}}\left( \left \vert 00 \right \rangle_{AB} + \left
\vert 11 \right \rangle_{AB} \right)$. Next, Alice performs a joint
measurement of the 2-qubit system $X \otimes A$ in the Bell basis:
\begin{eqnarray} 
\left \vert \Phi^{\pm} \right \rangle=\frac{1}{\sqrt{2}}\left (\left
\vert 00 \right \rangle \pm \left \vert 11 \right \rangle \right),\\
\left \vert \Psi^{\pm} \right \rangle =\frac{1}{\sqrt{2}}\left (\left
\vert 01 \right \rangle \pm \left \vert 10 \right \rangle \right).
\end{eqnarray} 
Finally, Alice sends a two-bit message to Bob informing him of her
measurement result. Bob then rotates his qubit using one of the
unitary operators {\bf 1}, $\sigma_z$, $\sigma_x$ or $\sigma_y$,
according to whether Alice's result was respectively $\ket{\Phi^{+}}$,
$\ket{\Phi^{-}}$, $\ket{\Psi^{+}}$ or $\ket{\Psi^{-}}$. The final
state of Bob's qubit is then equal to the original state
$\ket{\phi}_X$, regardless of the measurement result. This
insensitivity to measurement results is the crucial property of the
teleportation protocol, and one which we shall also require for our
telecloning scheme.

\subsection{Optimal Universal Quantum Cloning}
\label{sec:uqcm}
While teleportation aims to {\it transmit} quantum information
faithfully, optimal cloning seeks to {\it spread} it among several
parties in the most efficient way possible. The `no-cloning' theorem
\cite{Wootters} prevents this spreading from being perfect;
nevertheless, it is still reasonable to ask {\it how accurately} can
such copies be made \cite{Buzek0}. If the quality of the copies
(measured, for instance, by their fidelity with respect to the
original state $\ket{\phi_{in}}$) is chosen to be independent of
$\ket{\phi_{in}}$, then the answer is given by the so-called $N
\rightarrow M$ Universal Quantum Cloning Machines (UQCMs)
\cite{Gisin}.

These `machines' are unitary transformations that transform $N$ input
systems which are identically prepared in state $\ket{\phi_{in}}$ onto
$M$ output systems $\left( M \geq N \right)$, each of which ends up in
a mixed state described by the reduced density operator
\begin{eqnarray}
\rho_{out}= \gamma \ketbra{\phi_{in}}{\phi_{in}} + \left( 1- \gamma
\right) \ketbra{\phi_{in}^\perp}{\phi_{in}^\perp} 
\label{eqn:shrinking}
\end{eqnarray} 
(where $\ket{\phi_{in}^\perp}$ is a state orthogonal to
$\ket{\phi_{in}}$) \cite{Gisin,Buzek2}. The fidelity factor $\gamma$
of these imperfect copies has a definite upper limit imposed by
quantum mechanics. In the case where each input system consists of one
qubit, this optimal value is given by \cite{Gisin}:
\begin{eqnarray}
\gamma=\frac{M \left(N+1 \right)+N}{M
\left(N+2 \right)}.
\label{eqn:gamma}
\end{eqnarray} 
Unitary transformations which realize this bound have also been found
\cite{Gisin}. In general, they involve the $N$ `original' qubits,
$M-N$ `blank paper' qubits (initially prepared in some fixed state
$\ket{0 \cdots 0}_B$), and an ancilla system $A$ containing at least
$M-N+1$ levels (also initially in some fixed state $\ket{0 \cdots
0}_A$). In this paper, we shall be mainly interested in the situation
where only one original qubit is available, that is, $N=1$. In this
case, the cloning transformation $U_{1M}$ is defined as follows: for
an initial state $\ket{\phi_{in}}=a\ket{0}+b\ket{1}$, we have
\begin{eqnarray}
U_{1M}\left(
\ket{\phi_{in}}
\otimes \ket{0 \cdots 0}_A \ket{0 \cdots 0}_B 
\right) \nonumber \\
=a\ket{\phi_0}_{AC}+b\ket{\phi_1}_{AC},
\label{eqn:1mclon}
\end{eqnarray}
where 
\begin{eqnarray}
\ket{\phi_0}_{AC}&=&U_{1M}\ket{0} \ket{0 \cdots 0}_A \ket{0 \cdots 0}_B
\nonumber \\
&=&\sum_{j=0}^{M-1}\alpha_j
\ket{A_j}_A\otimes \ket{\left\{0,M-j\right\},\left\{1,j\right\}}_C,
\label{eqn:oqcstate0}
\\
\ket{\phi_1}_{AC}&=&U_{1M}\ket{1}
\ket{0 \cdots 0}_A \ket{0 \cdots 0}_B
\nonumber \\
&=&\sum_{j=0}^{M-1}\alpha_j
\ket{A_{M-1-j}}_A\otimes \ket{\left\{0,j\right\},\left\{1,M-j\right\}}_C,
\label{eqn:oqcstate1}
\\
\alpha_j&=&\sqrt{\frac{2\left(M-j\right)}{M\left(M+1\right)}}
\end{eqnarray}
and where C denotes the $M$ qubits holding the copies.  Here,
$\ket{A_j}_A$ are $M$ orthogonal normalized states of the ancilla and
$\ket{\left\{0,M-j \right\}, \left \{1,j \right\}}$ denotes the
symmetric and normalized state of $M$ qubits where $(M-j)$ of them are
in state $\ket{0}$ and $j$ are in the orthogonal state $\ket{1}$. For
example, for $M=3,~j=1$:
\begin{eqnarray}
\ket{\left\{0,2 \right\}, \left \{1,1
\right\}}=\frac{1}{\sqrt{3}}\left( \ket{001} + \ket{010} +\ket{100}
\right).
\end{eqnarray}

We note that, even though the minimum number of ancilla qubits
required to support the $M$ levels $\ket{A_j}_A$ is of the order of
$\log_{2} M$, these can be more conveniently represented as the
symmetrized states of $\left (M-1 \right)$ qubits \cite{Buzek1}:
\begin{eqnarray}
\ket{A_j}_A \equiv \ket{\left\{0,M-1-j\right\},\left\{j,1\right\}}_A.
\label{eqn:ancilla}
\end{eqnarray}
In this form, states $\ket{\phi_0}$ and $\ket{\phi_1}$ above become
$(2M-1)$-qubit states, obeying the following simple symmetries:
\end{multicols}
\begin{minipage}{6.54truein}
\begin{figure}[H]
  \begin{center}
    \leavevmode
    \epsfxsize=6.00truein
    \epsfbox{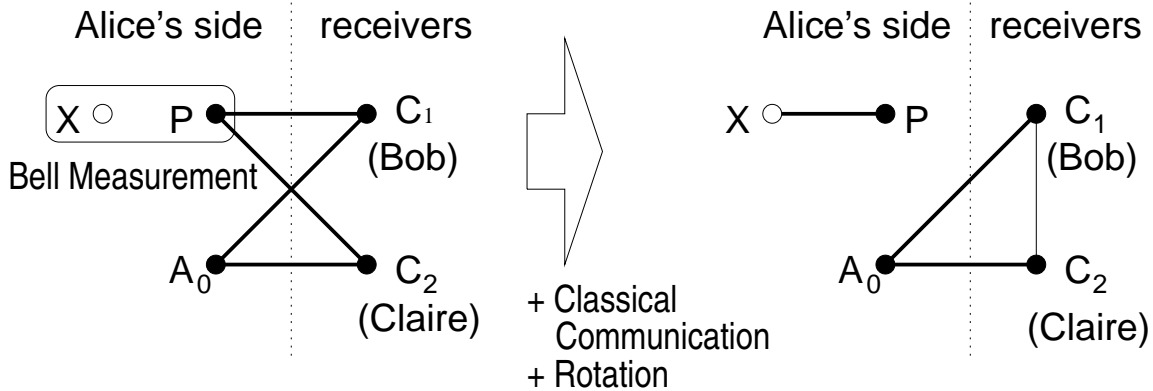}
  \end{center}
\caption{Quantum telecloning $M=2$ copies of an unknown 1-qubit
state. Alice and her associates Bob and Claire (the `receivers') 
initially share a multiparticle entangled state (Eq.~\ref{eqn:tcstate}) 
consisting of the qubits $P$ (the `port'), $A_0$ (the ancilla), $C_1$ 
and $C_2$ (outputs, or `copy' qubits). The solid lines indicate the 
existence of entanglement between pairs of qubits when the remaining 
ones are traced out. Alice performs a Bell measurement of the port 
along with the `input' qubit $X$; subsequently, the receivers perform 
appropriate rotations on the output qubits, obtaining two optimal 
quantum clones. Since these rotations are independent, each clone can 
be at a different location.}
\label{fig:tctwo}
\end{figure}
\end{minipage}
\begin{multicols}{2}
\begin{eqnarray}
\sigma_z\otimes \cdots \otimes\sigma_z \ket{\phi_0}&=&\ket{\phi_0}, 
\label{eqn:symm1}\\
\sigma_z\otimes \cdots \otimes\sigma_z \ket{\phi_1}&=&-\ket{\phi_1}, 
\label{eqn:symm2}\\
\sigma_x\otimes \cdots \otimes\sigma_x \ket{\phi_{0(1)}}&=&\ket{\phi_{1(0)}}.
\label{eqn:symm3}
\end{eqnarray}
In other words, the states $\ket{\phi_i}$ transform under simultaneous
action of the Pauli operators on all $(2M-1)$ qubits just as a single
qubit transforms under the corresponding single Pauli operator. We
also note that these operations are strictly local, that is, factorized 
into a product of independent rotations on each qubit. As we will see 
in the next section, these local symmetries play a crucial role, allowing 
cloning to be realized remotely via multiparticle entanglement.

\section{Quantum telecloning}
\label{sec:tc}

In this section, we present a `telecloning' scheme that combines cloning 
and teleportation. This is accomplished as follows: Alice holds an (unknown) 
1-qubit state $\ket{\phi}_X$ which she wishes to teleclone to M associates 
Bob, Claire, etc. We assume that they all share a multiparticle entangled 
state $\ket{\psi_{TC}}$ as a starting resource. This state must be chosen 
so that, after Alice performs a local measurement and informs the other 
parties of its result, the latter can each obtain an optimal copy given 
by Eq.~(\ref{eqn:shrinking}) using only local rotations.

A choice of $\ket{\psi_{TC}}$ with these properties is the following $2M$-qubit 
state:
\begin{eqnarray}
\left \vert \psi_{TC} \right \rangle = \left( \left | 0 \right
\rangle_P \otimes \left | \phi_0 \right \rangle_{AC} + \left | 1
\right \rangle_P \otimes \left | \phi_1 \right \rangle_{AC} \right)/
\sqrt{2},
\label{eqn:tcstate}
\end{eqnarray}
where $\left | \phi_0 \right \rangle_{AC}$ and $\left | \phi_1 \right
\rangle_{AC}$ are the optimal cloning states given by
Eqs.~(\ref{eqn:oqcstate0}) and (\ref{eqn:oqcstate1}).  Here, $C$
denotes the $M$ qubits which shall hold the copies, each of which is
held by one of Alice's associates. For convenience, we shall refer to
them collectively as `the receivers' (though it should be kept in mind
that they may all be far away from each other). $P$ represents a
single qubit held by Alice, which we shall refer to as the `port'
qubit.  Finally, $A$ denotes an $M-1$ qubit ancilla, which for
convenience we will also assume to be on Alice's side (even though,
once again, each qubit may in reality be at a different location).

The tensor product of $\ket{\psi_{TC}}$ with the unknown state $\left
\vert \phi \right \rangle_X= a \left \vert 0 \right \rangle_X+b \left
\vert 1 \right \rangle_X$ held by Alice is a $(2M+1)$-qubit
state. Rewriting it in a form that singles out the Bell basis of
qubits $X$ and $P$, we get
\begin{eqnarray}
\left \vert \psi \right \rangle_{XPAC}
&=& \left \vert \Phi^+ \right \rangle_{XP}
\left(a \left | \phi_0 \right \rangle_{AC}
+ b \left | \phi_1 \right \rangle_{AC} \right )/\sqrt{2} \nonumber \\
&+& \left \vert \Phi^- \right \rangle_{XP}
\left(a \left | \phi_0 \right \rangle_{AC}
- b \left | \phi_1 \right \rangle_{AC} \right )/\sqrt{2} \nonumber \\
&+& \left \vert \Psi^+ \right \rangle_{XP}
\left(b \left | \phi_0 \right \rangle_{AC}
+ a \left | \phi_1 \right \rangle_{AC} \right )/\sqrt{2} \nonumber \\
&+& \left \vert \Psi^- \right \rangle_{XP}
\left(b \left | \phi_0 \right \rangle_{AC}
- a \left | \phi_1 \right \rangle_{AC} \right)/\sqrt{2}.
\label{totalstate}
\end{eqnarray}
The telecloning of $\left \vert \phi \right \rangle_X$ can now be
accomplished by the following simple procedure:

\begin{enumerate}
\item{Alice performs a Bell measurement of qubits $X$ and $P$, obtaining
one of the four results $\ket{\Psi^{\pm}}_{XP}$,
$\ket{\Phi^{\pm}}_{XP}$. If the result is $\ket{\Phi^+}_{XP}$, then
subsystem $AC$ is projected precisely into the optimal cloning state
given in Eq.~(\ref{eqn:1mclon}). In this case, our task is
accomplished.}

\item{In case one of the other Bell states is obtained, we can still
recover the correct state of AC by exploiting the symmetries of states
$\left \vert \phi_0 \right \rangle_{AC}$ and $\left \vert \phi_1
\right \rangle_{AC}$ under the Pauli matrix operations
(Eqs.~(\ref{eqn:symm1}-\ref{eqn:symm3})). Specifically, if
$\ket{\Phi^-}_{XP}$ is obtained, we must perform $\sigma_z$ on each of
the $2M-1$ qubits in AC; similarly, if $\ket{\Psi^+}_{XP}$ or
$\ket{\Psi^-}_{XP}$ are obtained, they must all be rotated by
$\sigma_x$ and $\sigma_x\sigma_z=i\sigma_y$,
respectively.}
\end{enumerate}
This procedure is illustrated in Fig.~\ref{fig:tctwo} for the case of 
$M=2$ copies.

We stress that, apart from Alice's Bell measurement, only local
1-qubit operations are required in this telecloning procedure. In this
way, all of the qubits except the input $X$ and the port $P$ can be
spatially separated from each other. It is also worthwhile to add that
rotating the ancilla qubits in step (2) above is not strictly
necessary. The correct copy states of each output (given by
Eq.~(\ref{eqn:shrinking}) ) are obtained at the output regardless of
these operations, since local rotations on one qubit cannot affect
another qubit's reduced density operator.

We thus see that, given the telecloning state (\ref{eqn:tcstate}) and
using only local operations and classical communication, we are able
to optimally transfer information from one to several qubits. In the 
following section, we study in detail the entanglement properties of this 
state that allow this to happen.

\section{The Entanglement structure of the telecloning state}
\label{sec:ent}

The procedure we have described in the previous section performs the
same task as a unitary $1\rightarrow M$ cloning machine, but uses only
local operations and classical communication. In the former case,
information about the input state is conveyed to the output copies by
means of global entangling operations (this is explicitly shown in the
cloning network of Ref. \cite{Buzek1}). In telecloning, the same
transfer is realized through the multiparticle entanglement of state
(\ref{eqn:tcstate}).  In this section, we investigate the structure of
this entanglement. It is important to remark that at present there is
no known way of uniquely quantifying the entanglement of a general
multiparticle state \cite{Vedral}. For the purpose of understanding
the flow of information in the telecloning procedure, we find it
convenient to perform this analysis from two points of view, which we
refer to as the `total' and `two-qubit' pictures. The first of these
involves all $2M$ particles (hence `total'), and refers to the entanglement 
between the $M$ qubits on Alice's side (the port and ancilla) and the $M$ on
the receivers' side (the outputs); the second considers the
entanglement of a single pair of qubits after tracing over all other qubits.

Let us first consider the `total' picture. We begin by rewriting the
telecloning state (\ref{eqn:tcstate}) so that the qubits on Alice's
and the receivers' sides are explicitly separated:
\end{multicols}
\noindent\rule{0.5\textwidth}{0.4pt}\rule{0.4pt}{\baselineskip}
\begin{eqnarray}
\left \vert \psi_{TC} \right \rangle= \frac{1}{\sqrt{M+1}}
\sum_{j=0}^M{\left | \left \{ 0,M-j \right\}, \left \{ 1,j \right\}
\right \rangle_{PA}}
\otimes \left | \left \{ 0,M-j \right\}, \left 
\{ 1,j \right\} \right \rangle_C.
\label{eqn:tc}
\end{eqnarray}
\hspace*{\fill}\rule[0.4pt]{0.4pt}{\baselineskip}%
\rule[\baselineskip]{0.5\textwidth}{0.4pt}
\begin{multicols}{2}
This form highlights the high degree of symmetry of the telecloning
state: it is completely symmetric under the permutation of any two
particles on the same side, and also under the exchange of both sides. 
This implies that, in fact, any of the $2M$ qubits can be used as the
telecloning port, with the clones being created on the opposite side.
Another implication is that, instead of using all
$2^M$ levels of the $M$ qubits on each side, we only need to take into
account their $(M+1)$ symmetric states. These can be associated with
the states of an $(M+1)$-level particle as follows:
\begin{eqnarray}
\left | {\underline j} \right \rangle
\equiv
\left | \left \{ 0,M-j \right\}, \left \{
1,j \right\} \right \rangle.
\end{eqnarray}
(We note that this property arises from the choice of symmetric
ancilla states in Eq. (\ref{eqn:ancilla})). Noting the exchange symmetry
 of both sides of Eq. (\ref{eqn:tc}), the telecloning state
(\ref{eqn:tc}) can then be conveniently rewritten as the following {\it
maximally entangled} state of two $(M+1)$-level particles \cite{Cerf}
\begin{eqnarray}
\left \vert \psi_{TC} \right \rangle=
\frac{1}{\sqrt{M+1}}
\sum_{j=0}^M{\left | {\underline j} \right \rangle_{PA}} \otimes
\left | {\underline j} \right \rangle_C.
\end{eqnarray}
The corresponding amount of entanglement, given by the von Neumann
entropy of each side's reduced density operator, is
$\varepsilon(\ket{\psi_{TC}})=\log_2 \left(M+1 \right)$.

We now show that this is in fact the {\it minimum} amount
necessary for any telecloning scheme based on the cloning transformation 
defined by Eq. (\ref{eqn:1mclon}). To see this, suppose the input qubit 
$X$ is already maximally entangled with another qubit $D$
\begin{equation}
\ket{\phi_{in}} =\frac{1}{\sqrt{2}}\left( 
\ket{0}_D\ket{0}_X+\ket{1}_D\ket{1}_X \right).
\end{equation}
Then the linearity of transformation (\ref{eqn:1mclon}) implies that 
the output of the cloning procedure must be
\begin{equation}
\ket{\phi_{out}} =\frac{1}{\sqrt{2}}\left( \ket{0}_D\ket{\phi_0}_{AC}+
\ket{1}_D\ket{\phi_1}_{AC} \right),
\end{equation}
which is precisely our telecloning state $\ket{\psi_{TC}}$. Therefore,
a telecloning scheme where $AD$ and $C$ are spatially separated allows
the creation of at least $\log_2 \left(M+1\right)$ e-bits, between two
distant parties. We know however that entanglement cannot be increased
only by local operations and classical communication \cite{Vedral}. We
must conclude then that any telecloning scheme based on
Eq.~(\ref{eqn:1mclon}) requires at least $\log_2 \left(M+1\right)$
e-bits between these parties as an initial resource. The scheme we
have described above is therefore optimal in this sense.

In contrast, if Alice used a local unitary network to obtain $M$
clones, and then teleported each one separately to its recipient, the
amount of entanglement required would be $M$ e-bits. Thus, telecloning
realizes the same task with an much more efficient use of
entanglement. If we also consider, as noted above, that the
telecloning state allows any of the $2M$ qubits to function as a port,
then the increase in efficiency is even greater (in order to allow the
same freedom of choice in the `clone then teleport' protocol, $M^2$
singlets would be necessary). Of course, in the case where only one
`clone' is produced ($M=1$), the telecloning state is just a maximally
entangled state of two 2-level systems (in other words a Bell
state). In this case, our scheme reduces to the usual teleportation
protocol.

While entanglement between the two sides gives a measure of the
resources necessary to accomplish telecloning, the `two-qubit'
entanglement between an arbitrary pair of particles helps track how
information from Alice's unknown state is conveyed to the clones. To
see this, we first calculate the reduced density matrix of each pair
of qubits. Due to the symmetries of the telecloning state, there are
only two different classes of pairs: those where both qubits are on
opposite sides (Alice's and the receivers') and those where they are
on the same side.

In the first case, the reduced joint density matrix of the two qubits
in the $\left \{ \ket{00}, \ket{01}, \ket{10}, \ket{11} \right \}$
basis is 
\begin{eqnarray}
\rho_{PC}=\frac{1}{6M}\left(
\begin{array}{cccc}
2M+1&0&0&M+2\\
0&M-1&0&0\\
0&0&M-1&0\\
M+2&0&0&2M+1\\
\end{array}
\right).
\label{eqn:rdpc}
\end{eqnarray}
The Peres-Horodecki theorem \cite{Peres,Horodecki} provides us with a
simple algorithm for determining whether or not a general two-qubit
state is entangled. All that is necessary is to calculate the
eigenvalues of the partial transpose of the state's density
matrix. According to the theorem, a state is entangled if and only if
at least one of these eigenvalues is negative. The partial transpose
of Eq.~(\ref{eqn:rdpc}) is 
\begin{eqnarray}
\rho_{PC}^{T_2}=\frac{1}{6M}\left(
\begin{array}{cccc}
2M+1&0&0&0\\
0&M-1&M+2&0\\
0&M+2&M-1&0\\
0&0&0&2M+1\\
\end{array}
\right)
\label{eqn:transpose}
\end{eqnarray}
The smallest eigenvalue of this matrix is $-1/\left( 2M \right)$, so
that state $\rho_{PC}$ is always entangled for all $M$. Thus, any pair
of qubits on opposite sides of the telecloning state (in particular,
the ones used as port and outputs) will be entangled, and by the same
amount.  On the other hand, the reduced density matrix for two qubits
which are both on the same side is
\begin{eqnarray}
\rho_{PA}=\frac{1}{6}\left(
\begin{array}{cccc}
2&0&0&0\\
0&1&1&0\\
0&1&1&0\\
0&0&0&2\\
\end{array}
\right).
\end{eqnarray}
This reduced density matrix is independent of $M$, and the minimum
eigenvalue of its partial transpose is $1/6$. Thus, any two qubits on
the same side of the telecloning state are disentangled. However,
their von Neumann mutual information
\begin{eqnarray}
I_{vN}=2 \ln 2 + \frac{1}{3} \ln \frac{1}{54}=0.0817,
\end{eqnarray}
is nonzero, which indicates that the copies on the receivers' side are
still classically correlated, although these correlations are weak.

The particular structure of the telecloning state can be justified
qualitatively in the following way: first of all, we certainly expect
Alice's port qubit to be entangled with the outputs, since without
entanglement quantum information cannot be sent using only a classical
channel. In addition, since all clones should be equal, the state
should be symmetric under permutations of the output qubits; in
particular, they should all be equally entangled with the
port. Furthermore, in order to optimize the transfer of information
the entanglement of the receiving and transmitting sides should be as
large as possible. Since the clones are symmetrized, and therefore
occupy only $M+1$ levels of their Hilbert space, the Schmidt
decomposition then implies that the total `two-side' entanglement
should be precisely that of two maximally entangled $(M+1)$-level
particles.  Finally, since the ancilla states on Alice's side may be
freely chosen (as long as they are orthogonal), it is natural to
assume them to be symmetrized, so that both sides are invariant under
permutation.

\begin{minipage}{3.27truein}
\begin{figure}[H]
  \begin{center}
    \leavevmode
    \epsfxsize=5cm
    \epsfbox{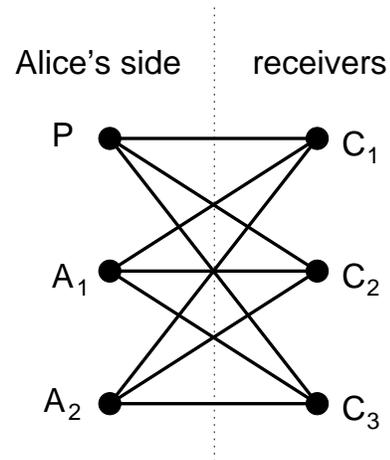}
  \end{center}
\caption{The telecloning state for $M=3$, consisting of one `port'
qubit, P two ancilla qubits (${\rm A}_1$ and ${\rm A}_2$) and three
output qubits (${\rm C}_{1-3}$). Solid lines indicate the existence of
two-qubit entanglement. Due to the symmetries of the state, the roles
of the port and ancilla qubits may be interchanged, as well as those
of the transmitting and receiving sides.}
\label{fig:tcthree}
\end{figure}
\smallskip
\end{minipage}

The calculations above also allow us to view the telecloning state as
a `network' of entangled qubits, each of which is only connected to
the $M$ ones on the opposite side (so the total number of `links' is
$M^2$; see Fig.~\ref{fig:tcthree}). 
Essentially, we may think of these
2-qubit connections as `communication channels' through which quantum
information may travel (in the same sense that Bennett {\it et al.}
referred to the Bell state in the original teleportation scheme as an
`EPR channel' \cite{Bennett0}). In this sense, the multiparticle
entanglement structure functions as a {\it multiuser channel},
allowing quantum information from Alice's input state to be conveyed
to all the output clones. This is emphasized by the fact that any
qubit in the network can be used as a port for the transmission.

\section{Open questions}
\label{sec:open}

Our work leaves a number of open questions which we now briefly
discuss. First of all, what is the most efficient way of generating
the telecloning state? In particular, we would like to find a way for
Alice and other users to create this state just by starting with
$\log(M+1)$ singlets and operating only locally with the aid of
classical communication. If Alice prepares the state locally and then
distributes the particles to other users these will in general travel
through a noisy channel. Then it would be important to find a
purification scheme to distill a ``good" telecloning state. The second
open question is whether our telecloning protocol is the most
efficient one or if there exists a way to use even less
entanglement. This might be possible if there exists a cloning
transformation which produces the same reduced density matrix for the
copies as in Eqs.~(\ref{eqn:shrinking}), (\ref{eqn:gamma}) but with
less entanglement between them and the ancilla. It is very important
to try to save on entanglement as much as we can, because this is the
resource that is hardest to manipulate and maintain in practice.  A
further task would be to generalize our scheme to telecloning of $N$
to $M$ particles.  Yet another generalization would be the telecloning
of $d$-dimensional registers \cite{Werner}.

\section{Summary}
\label{sec:summary}

We have presented a telecloning scheme which generalizes
teleportation by combining it with optimal quantum cloning. This
allows the optimal broadcasting of quantum information from one sender
(Alice) to $M$ spatially separated recipients, requiring only a single
measurement by Alice followed by classical communication and local
1-qubit rotations. Our scheme works by exploiting the multiparticle
entanglement structure of particular joint states of $2M$
particles. This structure can be seen as a multiuser `network'
connecting each qubit on Alice's side to each on the receivers' side,
in such a way that any node can be used to broadcast quantum
information to all those on the opposing side. The resulting state
requires only $\log_2 \left(M+1\right)$ e-bits of entanglement between
the two sides, representing a much more efficient use of entanglement
than the more straightforward approach where Alice first clones her
particle $M$ times and then uses $M$ singlets to transmit these states
to the different receivers.

In closing, we note that our scheme can also be applied to the
realization of a `quantum secret sharing' protocol as introduced
recently in \cite{Hillery}. This refers to the situation where Alice
wishes to teleport a 1-qubit state in such a way that it can only be
reconstructed at the `receiving' end of the teleportation channel if
two or more separate parties agree to collaborate. In our case, this
is accomplished by leaving both the ancilla and output qubits on the
receivers' side. Then Alice's original state may be reconstructed if
and only if all the output clones and ancilla qubits are brought
together to the the same location and acted upon by the inverse of the
cloning transformation $U_{1M}$ given in Eq.~(\ref{eqn:1mclon}).

\acknowledgements
The authors thank P.L. Knight for useful discussions. This work was
supported in part by the Japan Society for the Promotion of Science,
the Brazilian agency Conselho Nacional de Desenvolvimento
Cient\'{\i}fico e Tecnol\'{o}gico (CNPq), the Overseas Research
Student Award Scheme, the UK Engineering and Physical Sciences
Research Council, the Knight Trust, the Elsag-Bailey Company and the
European Community.

\end{multicols}

\end{document}